\documentclass[prd,nofootinbib,twocolumn]{revtex4}
\usepackage{graphics,epsfig}

\def\alm{a_{\ell m}}
\def\Cl{C_\ell}
\def\deg{^{\circ}}
\def\lmax{\ell_{\rm max}}

\begin{document}

\title{
Distribution of Singularities in the
Cosmic Microwave Background Polarization }

\author{
Dragan Huterer and Tanmay Vachaspati
}
\affiliation{
CERCA, Department of Physics, Case Western Reserve University,
10900 Euclid Avenue, Cleveland, OH 44106-7079, USA.}

\begin{abstract}
\noindent
The polarization of the cosmic microwave background radiation will
have a distribution of singularities and anti-singularities, points
where the polarization vanishes for topological reasons. The
statistics of polarization singularities provides a non-trivial scheme
to analyze the polarization maps that is distinct from the usual
two-point correlation functions. Here we characterize the statistics
of the singularity distribution in simulated polarization maps, and
make predictions that can be compared with ongoing and upcoming
observations. We use three different characterizations: the nearest
neighbor distance between singularities, the critical exponent $\nu$
that captures the scaling of total charge $q$ within a closed curve of
length $L$ ($q \propto L^{\nu}$), and the angular two point angular
correlation functions for singularities of similar and opposite charge.
In general, we find that the distribution of singularities is random
except on scales less than about $10\deg$, where singularities of the
same charge repel and those of opposite charge attract. These
conclusions appear to be extremely robust with respect to variations
in the underlying cosmological model and the presence of
non-Gaussianity; the only exception we found are cases where 
statistical isotropy is grossly violated. This suggests that, within
the assumption of statistical isotropy, the distribution is a robust
feature of the last scattering surface and potentially may be used as
a tool to discriminate effects that occur during photon propagation
from the last scattering surface to the present epoch.
\end{abstract}

\maketitle


The cosmic microwave background (CMB) anisotropies are linearly
polarized at the 10\% level. The polarization, predicted almost four
decades ago \cite{rees}, has recently been observed
\cite{dasi,wmap_pol} and efforts are underway to map it on
increasingly smaller scales. The polarization is most easily described
in terms of so-called E and B modes \cite{KamKosSte97,SeljZal_EB}. The
two-point correlation function of E and B maps is typically taken as a
statistic of choice to represent the polarization properties of the
map. The two-point function fully describes the map if it is
Gaussian random and isotropic, otherwise higher-order correlation
functions are necessary for the full description (for a 
review of CMB polarization, see e.g. Ref.~\cite{hu_white}).

\begin{figure}
\scalebox{0.40}{\includegraphics{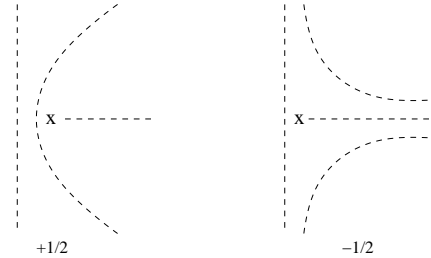}}
\caption{\label{fig:fund_def} Fundamental singularities of charge $\pm
1/2$. Each dash represents the linear polarization of the CMB
radiation at that point. The $\times$ marks the position of the
singularity where the linear polarization vanishes for topological
reasons.  }
\end{figure}

\begin{figure}[h]
\scalebox{0.40}{\includegraphics{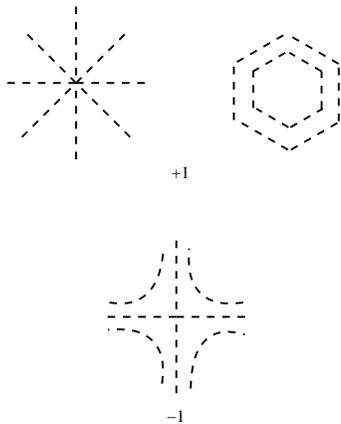}}
\caption{Singularities of charge
$\pm 1$, constructed by combining the two fundamental singularities
shown in
Fig.~\ref{fig:fund_def}. Top row shows examples of a ``knot'' and a
``focus'', while a ``saddle'' is shown at the bottom, using the
nomenclature from Ref.~\cite{NasNov98}.}
\label{fig:def_3kinds}
\end{figure}

In this paper we explore another, independent and largely
unexplored signature of CMB polarization: the distribution of
singularities \cite{NasNov98,dolgov,VacLue03}. The CMB polarization
map (denoted $P$) corresponds to a map from the sky -- the two
dimensional sphere, $S^2$ -- to the space of headless vectors, 
$S^2/Z_2$, given by the plane and amplitude of polarization:
\begin{equation}
P: S^2 \rightarrow S^2/ Z_2
\end{equation}
Such maps are known to contain topological features that are 
characterized by points with vanishing polarization, known as
``singularities'' or ``defects'', and each singularity carries a 
topological charge. The total topological charge within a closed
contour on the sky can be calculated as an integral over the
contour. This is very similar to Gauss' law that is used to determine
the electric charge within a closed surface by integrating the
electric field over the surface. The polarization singularities 
have fundamental charge $+1/2$. The anti-singularities have 
fundamental charge $-1/2$. These fundamental singularities are shown 
in Fig.~\ref{fig:fund_def}. They can be combined to form three kinds 
of double singularities as shown in Fig.~\ref{fig:def_3kinds}: 
``knots'' and ``foci'', which have charge $+1$, and ``saddles'' with 
charge $-1$ \cite{NasNov98}. The total charge in a given map is zero
for all practical purposes, as discussed in the next Section.

Polarization of the CMB, like the temperature, provides an extremely
important window to the processes in the early universe.  In fact, the
polarization has some advantages over the temperature, and in
particular it offers a more direct probe of the recombination era at
large angular scales \cite{hu_okamoto}. It is therefore well worthwhile to
consider alternative analyses of the polarization, and this motivates us
to explore the distribution of singularities. Our predictions can be
tested as soon as polarization maps become available. Furthermore,
since all of our statistics are computed in real (and not Fourier)
space, sky cuts due to galaxy and other foreground contamination will
be  relatively  easy to take into account.

\begin{figure*}[t]
\psfig{file=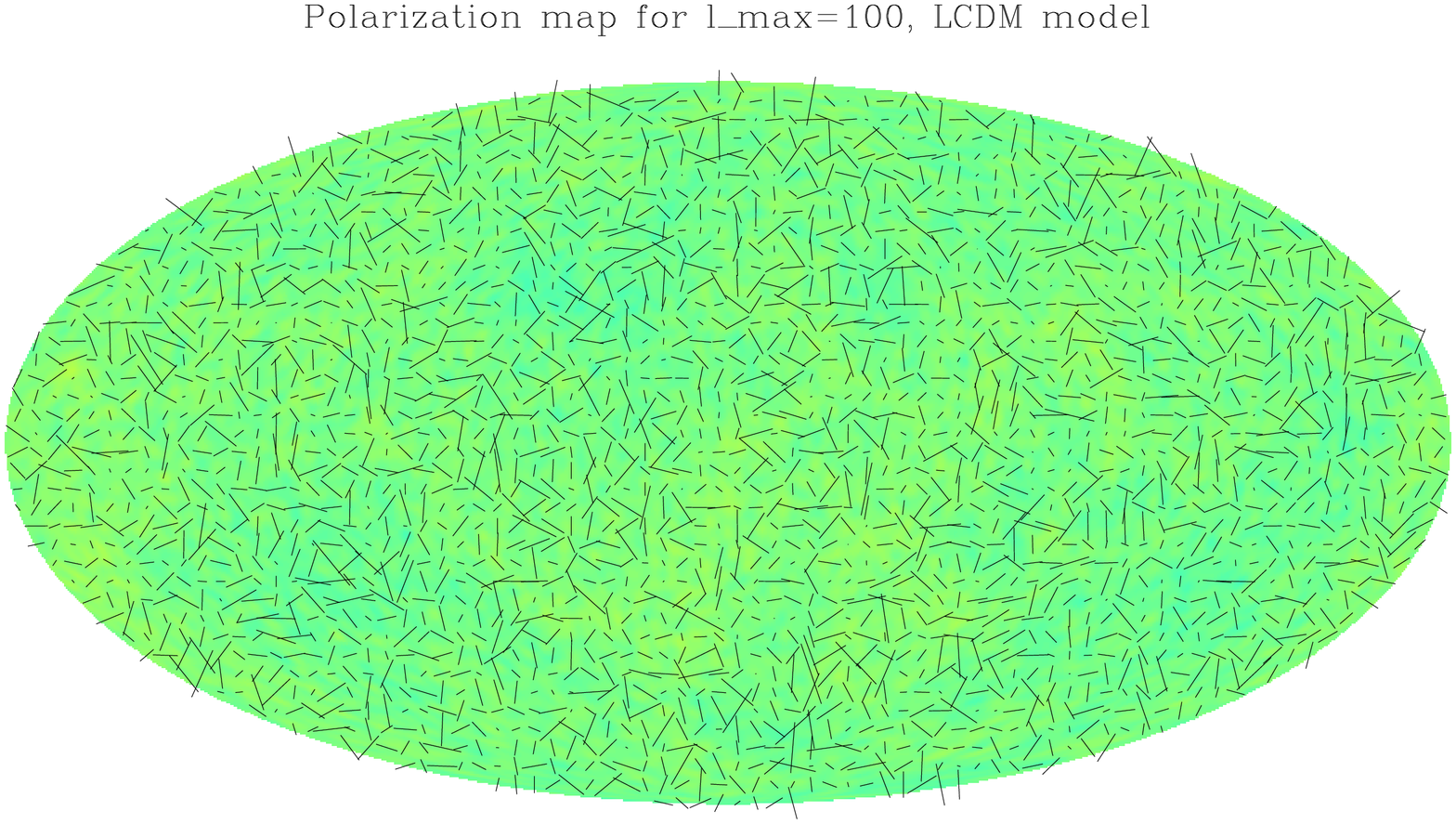,width=2.5in, height=3.5in, angle=90}
\psfig{file=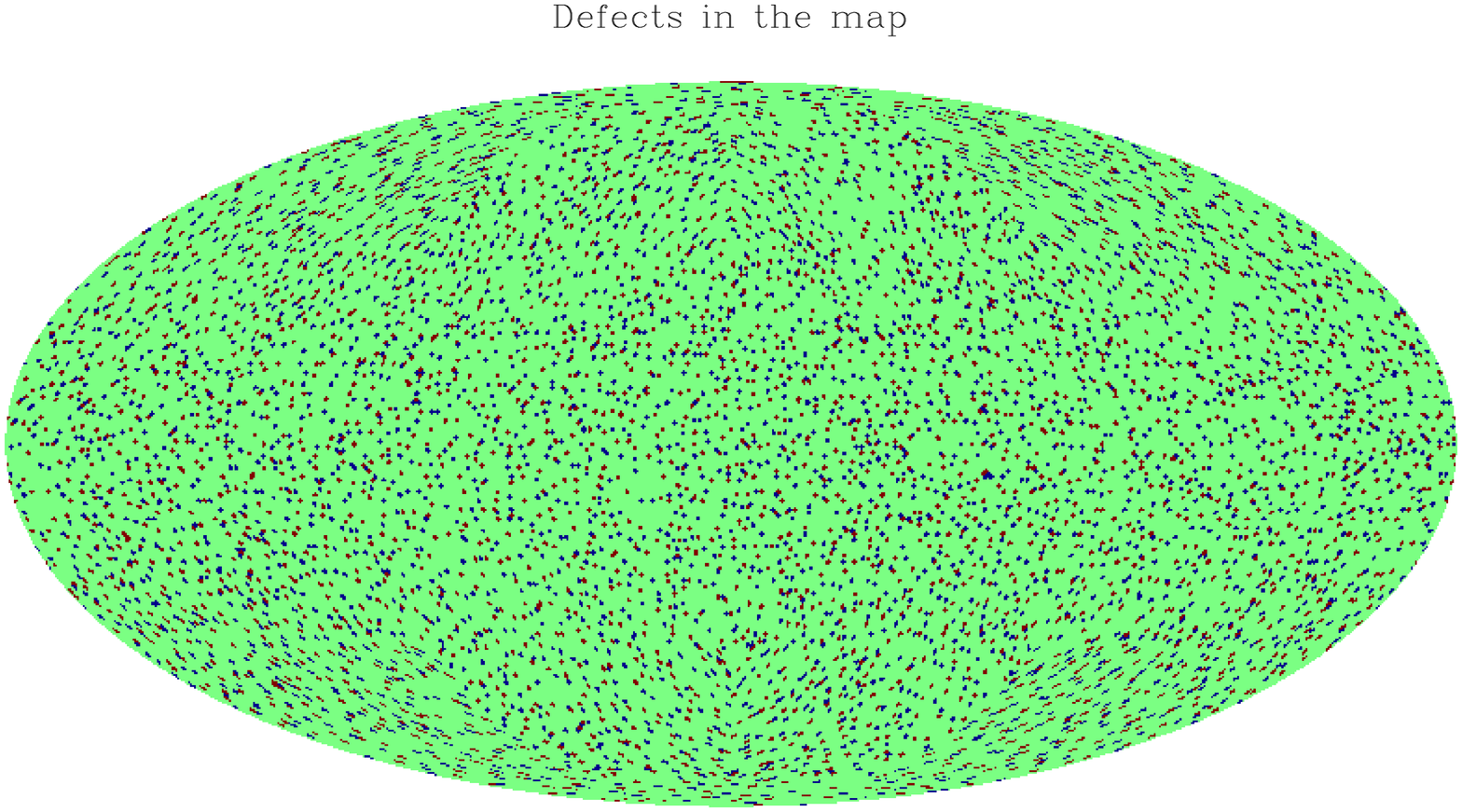,width=2.5in, height=3.5in, angle=90}
\caption{ {\it Left panel:} An example of the mock polarization
map. For visual clarity, this map was produced at the resolution
NSIDE$=256$, has power out to $\lmax=50$ and contains about 6000
singularities. {\it Right panel:} The distribution of singularities
in the same map. }
\label{fig:map}
\end{figure*}

Polarization singularities have first been described in
Refs.~\cite{NasNov98} and \cite{dolgov}. The expected distribution has
been described in Ref.~\cite{VacLue03} based on earlier experience
with condensed matter systems. We extend previous work and use a
highly quantitative analysis of the singularities, describe their
statistic as a function of pixelization scale and the underlying
cosmological model, and assess the statistical errors for each
measurement.

\section{The definition of singularities}\label{sec:definition}

The polarization patterns with fundamental
(charge $\pm 1/2$) singularities are shown in Fig.~\ref{fig:fund_def}.
The polarization map is described by the Stokes parameters
$Q$ and $U$. The properties of the polarization under rotations
imply that:
\begin{equation}
Q = I \cos (2 \alpha ) \ , \ \ U = I \sin (2 \alpha )
\end{equation}
where $I$ is the radiation intensity and $\alpha$ is the
polarization angle. The $\pm 1/2$ singularities are locations
around which $\alpha$ changes by $\pm \pi$. Therefore, given
a polarization map, we can find the change in $\alpha$ as we
go around a small closed path and this will tell us whether there
is a singularity within that closed path. By going around all possible
paths, we can find all the polarization singularities. In the
continuum, there would be an infinite number of small paths.
But in practice, the map is pixelized and the change in $\alpha$ 
is found around circuits defined by neighboring pixels.
The algorithm for finding the singularities in a pixelated
map is well-known and is described in the Appendix.

In this paper we will be concerned with the distribution of
polarization singularities. We will characterize the distribution in
three distinct ways. The first method is to find the total charge $q$
within a closed path of length $L$. The mean charge will, of course,
be zero because one can have positive and negative charges with equal
probability \footnote{Since the sky is a two sphere, the net charge is
+2. This is simply due to the topology of the two sphere -- it is
impossible to comb the hair on a two sphere without
singularities. This charge is a small number compared to the charges
that are present due to the statistical nature of the polarization map
and can be ignored for practical purposes.}. Next, we compute the
distribution of the distance to the nearest neighbor of a
singularity. Finally, motivated by the analyses of the galaxy
distribution on the sky, we compute the angular two-point correlation
function of singularities in the polarization map. This work
complements the pioneering theoretical discussions in
Refs.~\cite{NasNov98,dolgov,VacLue03} and helps establish the
distribution of singularities in the CMB polarization as a significant
probe of the universe.

In the next section, we describe how we create the mock polarization
maps. In Sec.~\ref{sec:stat} we find the critical exponents and the
angular correlation function, as well as the distribution of distance to the
nearest neighbor for a vanilla mock map. We investigate the effects of
non-random phases and non-Gaussianity in Sec.~\ref{sec:departures} and
summarize our results in Sec.~\ref{sec:conclusions}. In the Appendix 
we describe the algorithm for determining the presence of a singularity.

\section{Scheme to produce mock polarization maps}
\label{sec:mockmaps}

To produce mock polarization maps, we proceed as follows. We first
produce the angular power spectra (temperature and polarization) of
the CMB using the CMBFAST package \cite{cmbfast}.  Our fiducial model
is the standard $\Lambda$CDM cosmology with a flat universe with
matter energy density relative to critical $\Omega_M=0.3$, dark energy
equation of state $w=-1$, scalar spectral index $n_s=1.0$, physical
matter and baryon energy densities of $\Omega_M h^2=0.127$ and
$\Omega_B h^2=0.021$ respectively, and no tensor modes (we explore the
variations to this model in Sec.~\ref{sec:departures}). We can obtain
an arbitrary number of statistically independent maps by generating
different sets of coefficients $\alm^E$ and $\alm^B$, consistent with
the same underlying power spectra $\Cl^E$ and $\Cl^B$ respectively,
and generating the polarization map in Healpix \cite{healpix} for each
set separately.  Recall that the coefficients $\alm^{(E, B)}$ fully
describe a given map and, for a Gaussian random map, come from a
Gaussian distribution of variance $\Cl^{(E,B)}$. In some cases we will
want to produce polarization maps that do {\it not} come from Gaussian
random $\alm$, and in those cases we simply input the desired
non-Gaussian $\alm^{(E, B)}$ directly.

\section{Statistics of the Singularity Distribution}
\label{sec:stat}

\subsection{Number of singularities}\label{sec:Ndef}
   
We first compute the total number of singularities (or charges) $N_{\rm
sin}$, positive plus negative, in a given mock map. As expected, the
number of charges increases with the map resolution, just as, for
example, the number of temperature hot and cold spots increases with
resolution.  $N_{\rm sin}$ increases roughly as the square of the
maximum resolution of the map $\lmax$ so that, for example, a map
(consistent with our fiducial $\Lambda$CDM model) with $\lmax=100$ has
about 6000 singularities, while a map with $\lmax=500$ has about 150,000
singularities.  Note that, even if we are able to measure the polarization
with infinite resolution, $N_{\rm sin}$ does not increase
indefinitely, but levels off when the power in polarization becomes
negligible. For a standard $\Lambda$CDM model, the EE power spectrum
has power all the way to $\ell\sim 2000$. However, this implies that
the total number of singularities is likely to be larger than a million,
making the analysis of such a map challenging. Fortunately, we do not
need to worry about this issue, or wait for polarization experiments
that will reach scales this small, such as PLANCK, in order to explore
the distribution of singularities: the tests we propose can be performed for
polarization maps covering any range of angular scales, and 
statistics from the measured map of any given resolution can be
compared to Monte Carlo tests with mock maps of the same resolution.


\subsection{Scaling of RMS charge}\label{sec:qL}

We would now like to quantify how the {\it variance} in the number of
singularities increases with the area covered on any given map.  The 
root-mean-square (RMS) fluctuation of the charge within
a  closed path of length $L$ is expected to be:
\begin{equation}
\sigma(q(L))\equiv 
\left < \left (q(L)-\bar{q}(L)\right )^2\right >^{1/2} = a L^\nu
\end{equation}
where $q(L)$ is the total (positive plus negative) charge within $L$,
$\bar{q}(L)$ is its mean among the different paths of the same length,
$a$ is a system-dependent constant and the critical exponent $\nu$ is
expected to be 0.5 \cite{VacLue03}. Using our numerical analysis of
mock data we will be able to predict both $a$ and $\nu$ together with
error bars. 

To compute the RMS of charge per ring, we create a polarization map,
and pick a point on it that we call the North Pole. In the Healpix
representation of the map, each ring of angle $\theta$ from the North
Pole has length $L=2\pi\sin{\theta}$, and we find $q(L)$ for all rings
using the procedure described in the Appendix.  We then rotate the
north pole of the map in a random direction (i.e. assign it to a new,
randomly chosen point on the sphere) and repeat the computation of
$q(L)$ for each ring. We repeat this procedure hundred times in
order to obtain sufficient statistics and compute $\sigma(q(L))$.  The
mean charge per ring, averaged over all rotations, is nearly zero,
while the fluctuations around the mean is what we are interested in.

\begin{figure}
\centerline{\psfig{file=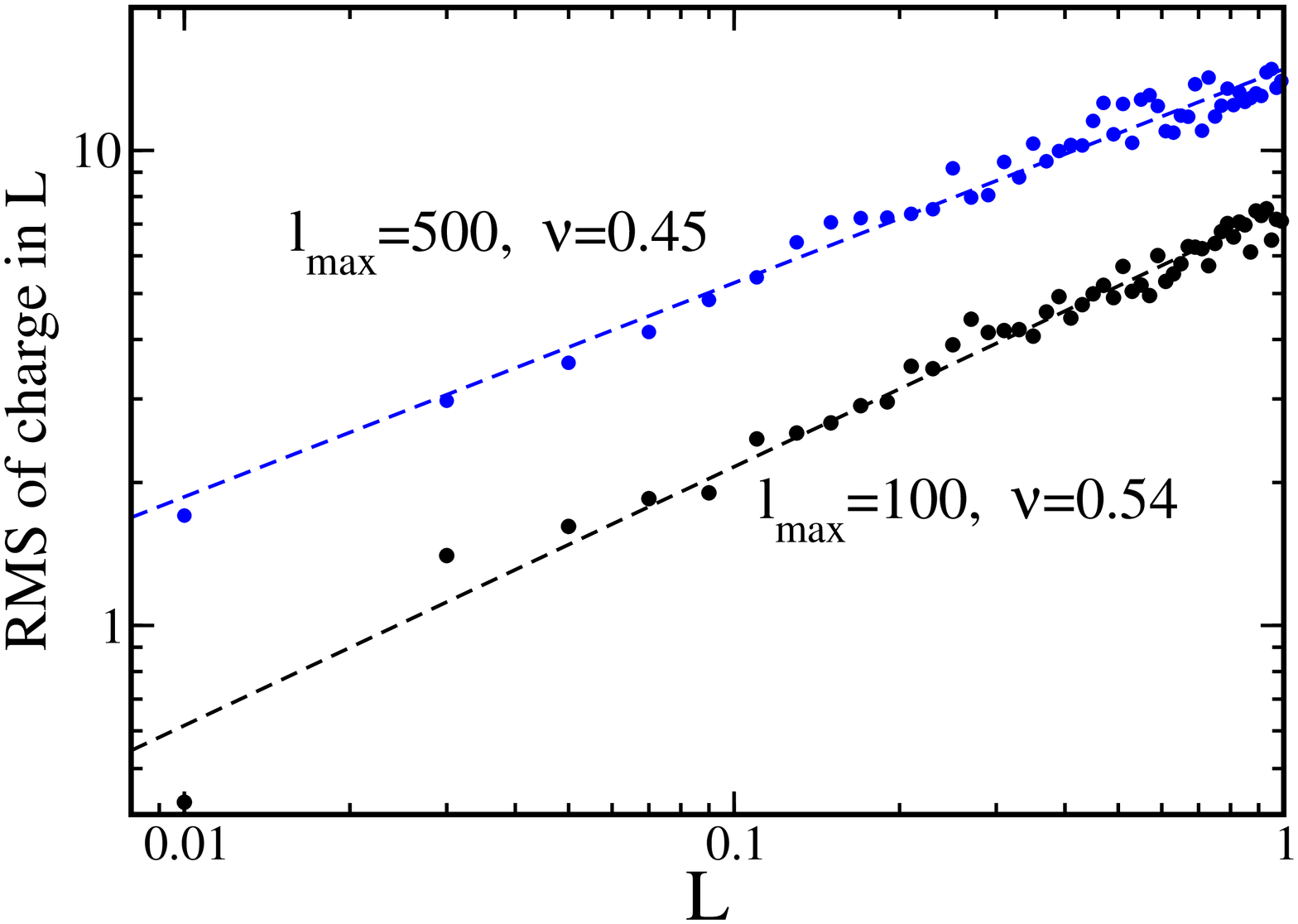,width=3.0in, height=3.8in, angle=-90}}
\vspace{-0.3cm}
\centerline{\psfig{file=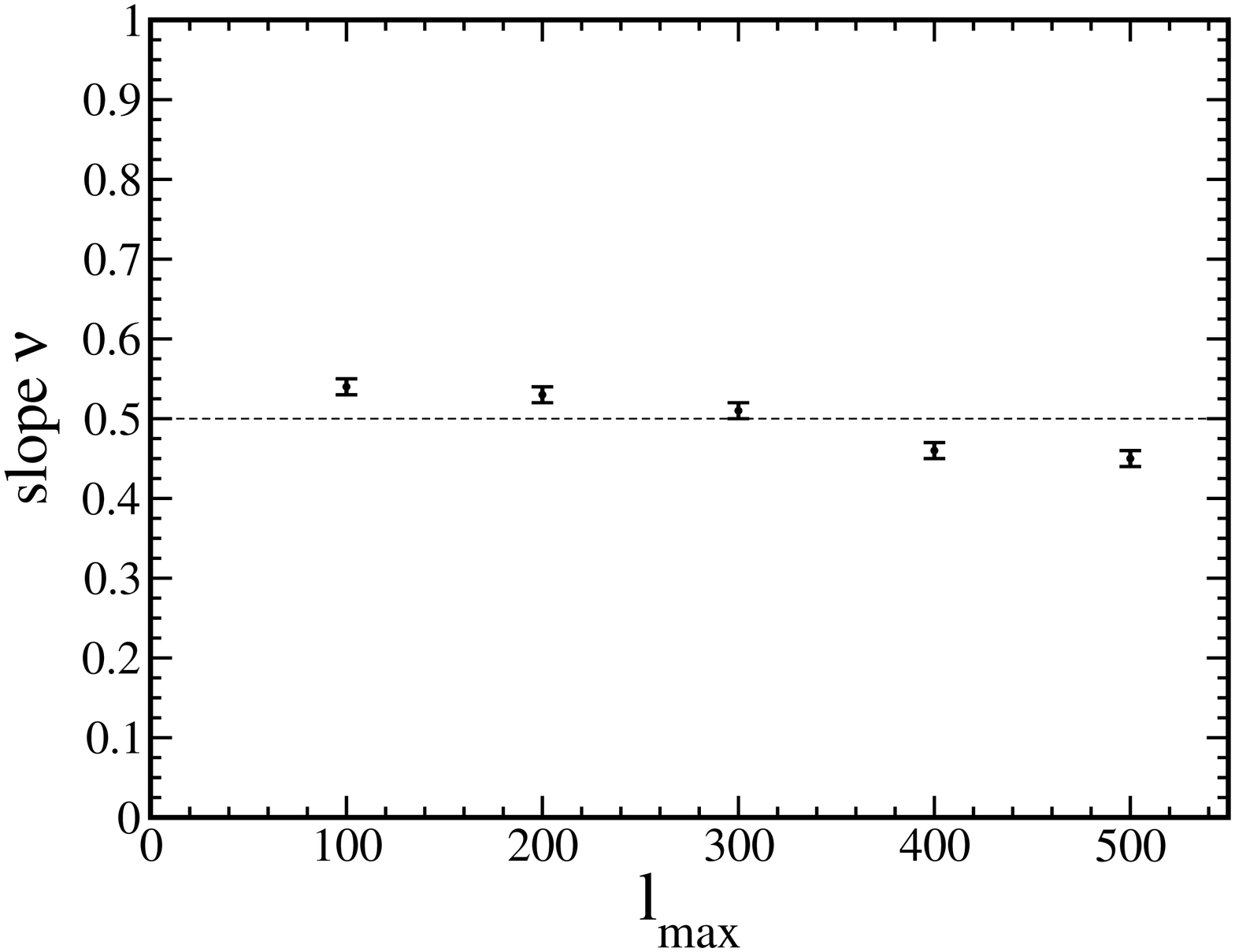,width=2.9in, height=3.8in, angle=-90}}
\caption{{\it Top panel:} Root-mean-square of the total charge per
ring as a function of the ring's length $L$.  The two datasets
correspond to polarization maps with resolution $\lmax=100$ (bottom)
and $\ell_{\rm max}=500$ (top), while the dashed curves denote the
linear fit in  the log-log coordinates. The distributions are
consistent with $\sigma(q(L))\propto L^{\nu}$ with $0.45\lesssim \nu
\lesssim 0.55$. {\it Bottom panel:} the dependence of $\nu$ on the
maximum resolution of the map $\lmax$ ($\lmax$ is the maximum
multipole where the map has power). }
\label{fig:critexp}
\end{figure}

Figure \ref{fig:critexp} (top panel) shows the scaling of
$\sigma(q(L))$ with $L$ for maps of two resolutions $\lmax=100$ and
$500$, corresponding to polarization having power down to scales of
$\sim 1\deg$ and $\sim 0.4\deg$ respectively. In both cases we have
fixed the pixelization of the map to the Healpix parameter
NSIDE$=256$, corresponding to pixels of about $0.25\deg$ on a side.
In these and many other cases we have explored, the RMS of charge
scales as $L^\nu$ where the critical exponent $\nu$ is nearly 0.5, in
agreement with expectations.  Furthermore, the exponent is independent
of the map pixelization NSIDE\footnote{ It is important to use NSIDE
large enough to capture the resolution of the map and avoid pixel
effects. This corresponds to NSIDE$>\lmax/4$.}. However, we find that
the exponent slightly decreases with map resolution, moving from
$\approx 0.55$ for $\lmax=100$ to $\approx 0.45$ for $\lmax=500$; see
the bottom panel of Fig.~\ref{fig:critexp}. In all cases the error in
the exponent is about $0.01$.

\subsection{Distance to the nearest neighbor}\label{sec:nn}

\begin{figure}
\centerline{\psfig{file=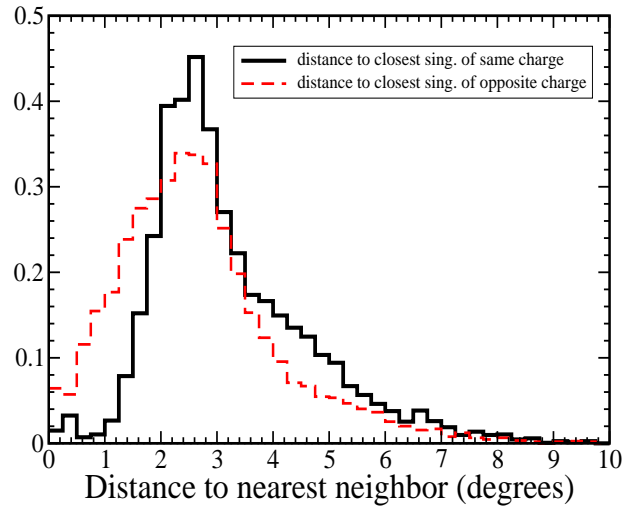,width=3.2in, height=3.8in, angle=-90}}
\caption{ Histogram of the distance from any given singularity to the 
nearest neighbor of the same charge (black-solid) and opposite charge
(red-dashed). Note the effects of repulsion between charges of the 
same sign, and attraction of charges of the opposite sign.}
\label{fig:nn}
\end{figure}

Another statistic that we explore is the distance to the nearest
neighbor from any given singularity. The histogram of the distances to
the nearest neighbor is shown in Fig.~\ref{fig:nn}, where for
computational convenience we have assumed a map with $\lmax=100$ which
has a total of about $6000$ singularities.  The average distance to
the nearest neighbor is slightly above $2\deg$ and can be roughly
predicted from the total number of singularities ($\sim 6000$
singularities in $\sim 40000$ degrees on the sky).  However, the
histograms of the nearest neighbor of the same charge and that of the
opposite charge are different, and show that opposite charges attract
and similar charges repel. For example, the mean distance to the
nearest neighbor of the same charge is $(3.24 \pm 0.01)\deg$, while
distance to the neighbor of the opposite charge is $(2.69 \pm
0.01)\deg$.

\subsection{Angular power spectrum of singularities}
\label{sec:wtheta}

The third characterization of the singularity distribution we suggest
and explore is the two point angular correlation function
$w(\theta)$. The angular two point function is simply the excess
probability, on top of expectation due to random distribution, of
finding one singularity at an angular location ${\vec{\theta}}_1$ 
and another one at location ${\vec{\theta}}_2$

\begin{eqnarray}
dP(\vec{\theta}_1, \vec{\theta}_2) 
&=& dP(\vec{\theta}_1)\, 
dP(\vec{\theta}_2)\, (1+w(\vec{\theta}_1-\vec{\theta_2}))\\
&=&dP(\vec{\theta}_1)\, dP(\vec{\theta}_2)\, (1+w(\theta))
\label{eq:wtheta} 
\end{eqnarray}

\noindent where $\theta =|\,\vec{\theta}_1-\vec{\theta_2}|$ and the
second line assumes statistical isotropy. In addition to the angular
correlation function for all singularities, we can find individual $w(\theta)$
for singularities of positive (or negative) charge in a similar way.

The angular correlation function is one of the standard tools to
describe the angular clustering of galaxies, and has been thoroughly
explored and used during the past three decades.  Computing
$w(\theta)$ for the galaxy distribution, however, typically involves
various practical problems, most important of which is the ``selection
function'' of the survey, having to do with magnitude cuts and
imperfect coverage across the field of view. The application we are
considering here is vastly simpler, since the singularities are
discrete, well-defined and easily computable features, and unlike
the galaxies they are located at the same radial distance. Furthermore, 
we are simulating the polarization pattern on full skies and do not
need to worry about edge-effects due to incomplete sky coverage. Therefore,
a simple estimator for $w(\theta)$ will suffice. We adopt the
Peebles-Hauser \cite{PeeHau} estimator
\begin{equation}
w(\theta)={N_{\rm rand}\over N_{\rm  map}} 
\left ({DD(\theta)\over RR(\theta)}-1 \right )
\end{equation}
\noindent where $DD(\theta)$ is the number of singularity pairs
separated by a distance larger than $\theta-d\theta$ but smaller than
$\theta+d\theta$, $RR(\theta)$ is the number of pairs from a random
(Poisson-distributed) map being in the same distance interval, and
$N_{\rm map}$ and $N_{\rm rand}$ are the total numbers of
singularities in the two maps.  In other words, the angular
correlation function measures the excess clustering over that
predicted by random distribution on any given scale. Here we use
$d\theta=5^{\circ}$, which is larger than our pixelization scale and
thus avoids any edge-effects due to pixelization. We have tried
several other values of $d\theta$ and obtained consistent results.

\begin{figure}
\centerline{\psfig{file=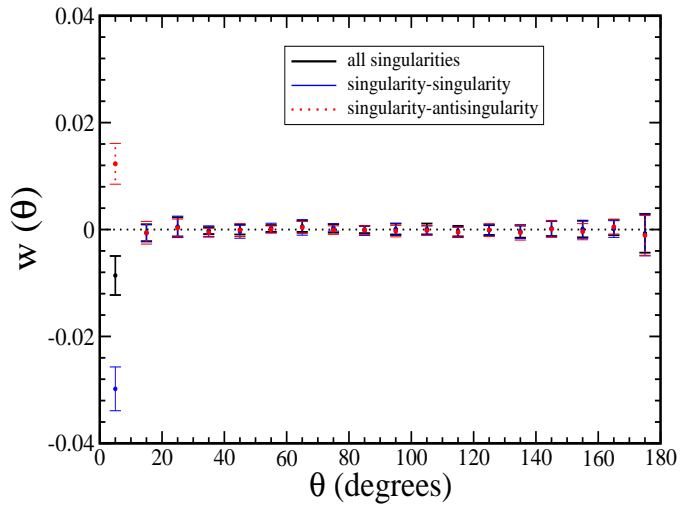, width=3.2in,
    height=3.8in, angle=-90}}
\caption{Angular correlation function of the distribution of
singularities, showing the excess probability of clustering over that
predicted by the random distribution.  The overall distribution of
singularities is consistent with random, except at small scales ($\lesssim
10\deg$) where singularities of opposite charge attract while those of
equal charge repel.}
\label{fig:wtheta}
\end{figure}

Figure \ref{fig:wtheta} shows the angular two-point correlation
function for the map with $\lmax=100$.  In each case we compute
$w(\theta)$ for 10 statistically independent maps, and plot the mean
and standard deviation of measurements at all angular scales. Note
that this procedure correctly accounts for the increase in the error
bars at large angular separations due to sample (or cosmic) variance.  

We find that $w(\theta)$ is remarkably consistent with zero at all
angular scales greater than about $10\deg$! This implies that the
singularities are Poisson distributed over most angular scales.  Also
note that the singularity-singularity $w(\theta)$ is negative at
scales $\lesssim 10\deg$, while the singularity-antisingularity
$w(\theta)$ is positive over the same range. This illustrates the fact
we discussed in Sec.~\ref{sec:nn}, that charges of the same sign
repel and those of opposite sign attract. Finally, $w(\theta)$ for
all singularities (irrespective of their charge) shows an overall
decrease at $\theta\lesssim 10\deg$.  We have repeated this analysis
by varying the maximum resolution of the map $\lmax$ and found
consistent results: the behavior of $w(\theta)$ is qualitatively
similar, while the error bars decrease with increasing $\lmax$ due to
the increase in the number of singularities.

\section{Exploring departures from the standard cosmological model}
\label{sec:departures}

It is important to determine how the singularity distribution depends
on the physical input such as the primordial fluctuation spectrum, the
cosmological parameters, or Gaussianity of cosmological seed
perturbations. In particular, previous work \cite{NasNov98,dolgov} 
emphasized that the distribution and type of singularities may be
a promising way of probing the Gaussianity of initial conditions.

With this in mind we have created mock maps of the CMB polarization
using several different cosmological models and characterized the
singularity distribution in each.  In particular, we have tried
several extreme possibilities, some of which are already ruled out by
cosmological observations:

\begin{itemize}
\item Maps based on primordial power spectrum with very significant
tensor modes (the ratio of tensor to scalar perturbations at
the CMB temperature quadrupole of $T/S=10$).

\item Maps with a strongly tilted primordial power spectrum with
either less or more power on small scales: we alternatively assumed a
scalar spectral index $n_s=0$ or $n_s=2$.

\item Maps that are strongly non-Gaussian: we assumed the real and
imaginary parts of the coefficients
$\alm^{(E,B)}$ to be sampled from an exponential distribution
while keeping zero mean and variance equal to $\Cl^{(E,B)}/2$.
In other words, we have adopted a highly skewed distribution of the
coefficients $\alm^{(E,B)}$.

\end{itemize}

Remarkably we find that all statistics we considered are largely
unaffected. The critical exponent is unchanged within the errors, and
so is its dependence on $\lmax$.  The two point correlation function
$w(\theta)$ is also unchanged, being consistent with zero except on
scales less than $10\deg$.  The total number of singularities does change
for the alternative cosmological models, which is to be expected since
the total power in the map is affected each time. However, statistics
of the distribution of singularities are unchanged. 

In particular, we find it very interesting that the results are
insensitive to the model for non-Gaussianity we assumed. We have
explored this further for several other classes of variations to the
standard Gaussian/isotropic assumption, and found deviations from the
results described in Sec.~\ref{sec:stat} {\it only} in cases where the
statistics of the map were modified enough that the statistical
isotropy was grossly violated.  For example, maps where the only
nonzero $\alm^{(E,B)}$ coefficients were those with $m=0$ showed
deviation from results in Sec.~\ref{sec:stat}; however, inspection of
polarization maps in these cases indicate that such modifications lead
to huge violations of isotropy in the map that are easily detectable
with almost any reasonable statistical test. Conversely, more subtle
modifications of the power spectra (for example, setting the
quadrupole of temperature and polarization to zero) produced the same
results as our fiducial model.  Therefore the characteristics of the
singularity distribution seem to be robust features that are
insensitive to the cosmological inputs at the last scattering surface.

The robustness of the singularity distribution holds advantages as
well as disadvantages. One disadvantage is that by examining the
distribution of the singularities in the actual data, it is unlikely
that we will uncover something about the physics of the primordial
fluctuations (unless statistical isotropy is violated).  The advantage
is that since the distribution is robust, any distortions in it must
come during propagation of the photons from the last scattering
surface to us. Hence the distribution can be used as a probe of
cosmology at redshifts smaller than $1000$. For example, weak
gravitational lensing could distort the Poissonian nature of the
angular correlation functions; however, this effect will operate only
at small angular scales, as lensing distortions are typically a few
arcminutes.

\section{Conclusions}
\label{sec:conclusions}

We have explored the statistics of the distribution of singularities
in the CMB polarization maps. The existence of singularities are a
generic prediction for a headless vector field on a sphere. Their
distribution, however, has not been explored in the past except for
some generic scaling arguments. Here we have provided a quantitative
analysis of the distribution of singularities with positive and
negative charge, and argued that the singularities provide an
additional probe of the conditions at last scattering, largely
independent of the usual two-point correlation functions of
temperature.

We found that the singularities are distributed randomly at scales $\gtrsim
10\deg$.  The angular two-point correlation function of singularities 
vanishes at those scales, while the total charge within a closed path
of length $L$ scales as $L^\nu$ with $\nu\simeq 0.5$. On scales
$\lesssim 10\deg$, however, charges of the same sign repel while those
of opposite charge attract. The attraction and repulsion are
manifested both in the angular two-point correlation function and in
the distribution of the nearest-neighbor distance.  

Perhaps surprisingly, we found that the aforementioned results are
very robust with respect to variations in the underlying cosmological
model assumed to create the mock maps. Changes in the tensor to scalar
ratio, scalar spectral index, and non-Gaussian distributions of the
$a_{lm}$'s all leave the singularity distribution unchanged. Only
violations of statistical isotropy affected the distribution of
singularities.  It is still possible that there are some other forms
of non-Gaussianity that can affect the singularity distribution that
we have not explored.  While this is impossible to rule out, it seems
more likely to us that the distribution of singularities at the last
scattering surface is described precisely by the Poissonian
distribution and other characteristics we have found. This implies
that any observed deviations of the actual map from these
distributions will have to be due to line of sight effects. Most
importantly, gravitational lensing of the CMB by the large-scale
structure can cause changes in the singularity distribution. However,
the lensing operates mostly on small scales, with typical deflections
of a few arcminutes and coherence of $< 10\deg$. Therefore, lensing
may affect the statistics of the singularity distribution only at
small scales.  We hope to explore this signature in future work.

\vspace{1cm}
\begin{acknowledgments} 
We thank Craig Copi for discussions.  This work was supported by the
U.S. Department of Energy. We have benefited from using the publicly
available CMBFAST \cite{cmbfast} and Healpix \cite{healpix}   packages.
\end{acknowledgments}

\appendix
\section{Calculation of the winding number}
\label{appendix}

Here we will describe how the winding of polarization around a 
contour $\Gamma$, {\it i.e.} net topological charge within $\Gamma$, 
is calculated.

The polarization map specifies $Q (\theta, \phi)$ and $U (\theta , \phi)$, 
where $\theta$ and $\phi$ are galactic latitude and longitude. From 
$Q$ and $U$ we can determine the polarization angle $\alpha$:
\begin{equation}
\alpha (\theta , \phi ) = \frac{1}{2} {\tan}^{-1} \biggl (\frac{U}{Q} \biggr )
\label{alpha}
\end{equation}
Now we want to know the change in $\alpha$ as we go around the closed
contour $\Gamma$.

\begin{figure}
\scalebox{0.40}{\includegraphics{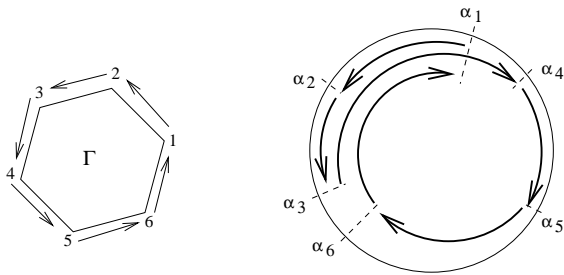}}
\caption{\label{fig:winding} To calculate the topological winding
along a contour $\Gamma$ -- here shown as a hexagon -- we follow
the phase $\alpha$, always choosing the shortest path on the
$\alpha$ circle. $\alpha_i$ denotes the value of the phase 
at point $i$ on $\Gamma$. The net change in $\alpha /2 \pi$ as
$\Gamma$ is traversed is the winding.
In the figure, the change in $\alpha$ is $- \pi$ since we go around
the $\alpha$ circle once in the counter-clockwise direction. Note 
that the full circle in $\alpha$ corresponds to an angle of only 
$\pi$, not $2\pi$. Hence the topological charge within $\Gamma$ in
the drawn example is $-1/2$.
}
\end{figure}

Any map of the CMB polarization will be pixelized. Hence only
an average value of $\alpha$ within each pixel will be available to
us. Then the change in $\alpha$ in going from pixel $i$ to a
neighboring pixel $i+1$ is given by:
\begin{equation}
\delta \alpha_i = \alpha_{i+1}-\alpha_{i} + \beta
\label{deltaalpha}
\end{equation}
where $\beta$ is defined as follows:
\begin{equation}
\beta =0 \ , \ \ \ {\rm if} \ |\alpha_{i+1}-\alpha_{i}| \le \pi /2 
\end{equation}
\begin{equation}
\beta =+\pi \ , \ \ \ {\rm if} \ \alpha_{i+1}-\alpha_{i} < - \pi /2 
\end{equation}
\begin{equation}
\beta =-\pi \ , \ \ \ {\rm if} \ \alpha_{i+1}-\alpha_{i} > + \pi /2 
\end{equation}
In other words $\delta \alpha$ is the shortest path from 
$\alpha_i$ to $\alpha_{i+1}$ around the circle defined by
$\alpha \in [-\pi /2, + \pi/2]$ (see Fig.~\ref{fig:winding}). Note that 
$\alpha = -\pi /2$ yields the same $Q$ and $U$ as $\alpha = +\pi /2$ 
and hence these two points are identified. 
 
The winding, $\Delta \alpha$, around the contour $\Gamma$ is now simply:
\begin{equation}
\Delta \alpha = \sum_i \delta \alpha_i
\end{equation}
where the sum is over all the discretized steps that define $\Gamma$.
The resulting topological charge is then $\Delta\alpha/(2\pi)$.

This scheme to find the windings has one small modification since
the sky is $S^2$. In this case, since the polarization
angle $\alpha$ is defined with respect to the lines of latitude
and this definition breaks down at the North and South poles,
we need to modify the scheme for any contour that contains the
North (or the South) pole. In that case, 1 must be added (per 
enclosed pole) to the net topological charge within the contour. 
This also ensures that the total topological charge on the sky is
$+2$ \cite{VacLue03}.

\end{document}